\documentstyle[preprint,12pt,aps,epsf]{revtex}

\begin{document}
\title{Evidence of strong antiferromagnetic coupling between localized and itinerant electrons in ferromagnetic $Sr_{2}FeMoO_{6}$ }
\author{M. Tovar, M.T. Causa, and A. Butera} 
\address{ Centro At\'{o}mico Bariloche and Instituto Balseiro\\Comisi\'{o}n 
Nacional de Energ\'{\i}a 
At\'{o}mica and Universidad Nacional de Cuyo,\\ 8400 
San Carlos de Bariloche, R\'{\i}o Negro, Argentina}
\author {J. Navarro, B. Mart\'{\i}nez, and J. Fontcuberta}
\address { Institut de Ci\`{e}ncia de Materials de Barcelona, Consejo 
Superior de Investigaci\'{o}n Cient\'{\i}fica, E-08193 Bellaterra, 
Catalunya, Spain}

\author { M. C. G. Passeggi }
\address{ Instituto de Desarrollo Tecnol\'{o}gico, Consejo Nacional de 
Investigaciones Cient\'{\i}ficas y T\'{e}cnicas and Facultad de 
Bioqu\'{\i}mica y Ciencias Biol\'{o}gicas, Universidad Nacional del 
Litoral,\\ 3000 Santa Fe, Santa Fe, Argentina }
\date{\today}
\maketitle

\begin{abstract}
Magnetic dc 
susceptibility ($\chi$) and electron spin resonance (ESR) measurements 
in the paramagnetic regime, are presented.  We found a Curie-Weiss (CW) behavior for 
$\chi$(T) with a ferromagnetic $\Theta = 446(5)$ K and 
$\mu_{eff} = 4.72(9) \mu_{B}/f.u.$, this being lower than that expected for either $Fe^{3+}$($5.9\mu_{B}$) or $Fe^{2+}$($4.9\mu_{B}$) 
ions.  The ESR $g$-factor  
$g = 2.01(2)$, is associated with $Fe^{3+}$.
We obtained an excellent description of the experiments in terms 
of two interacting sublattices: the localized $Fe^{3+}$ ($3d^{5}$) cores 
and the delocalized electrons.  The 
coupled equations were solved in a mean-field approximation, assuming for the 
itinerant 
electrons a bare susceptibility independent on $T$.  We 
obtained $\chi_{e}^{0} = 3.7$ $10^{-4}$ emu/mol. We show that the reduction of $\mu_{eff}$ 
for $Fe^{3+}$ arises from the strong antiferromagnetic (AFM) 
interaction between the two sublattices.  At 
variance with classical ferrimagnets, we found that $\Theta$ is 
ferromagnetic.  
Within the same model, we show that the ESR spectrum can be described 
by Bloch-Hasegawa type equations.  Bottleneck is 
evidenced by the absence of a $g$-shift. 
Surprisingly, as observed in CMR manganites, no narrowing effects of 
the ESR linewidth is detected in spite of the presence of the strong magnetic 
coupling.  These results provide evidence that 
the magnetic order in $Sr_{2}FeMoO_{6}$ does not 
originates in superexchange interactions, but from a novel mechanism recently proposed for double perovskites.

PACS:\ 75.10.-b, 76.30.-v, 76.60.Es, 75.30.Vn
\end{abstract}

\smallskip \newpage

\smallskip
The double perovskite $Sr_{2}FeMoO_{6}$ is known as a conducting ferromagnet 
(or ferrimagnet) with a relatively high transition temperature, 
$T_{c} > 400 K$, being magnetoresistant at room 
temperature.$^{1}$  The structure of $Sr_{2}FeMoO_{6}$ is built of 
perovskite blocks where the transition metal sites are alternatively occupied 
by $Fe$ and $Mo$ ions.  In the simplest ionic picture, the $Fe^{3+}(3d^{5}, S=5/2)$ 
ions was assumed to be antiferromagnetically (AFM) coupled to their six $Mo^{5+}(4d^{1}, S=1/2)$ 
nearest neighbors, leading to a total saturation magnetization, $M_{S} = 4\mu_{B}/f.u$.  
An alternative ionic description, giving the same $M_{S}$ value, assigned 
$Fe^{2+}(3d^{6}, S=2)$ and $Mo^{6+}(4d^{0})$, and assumed a ferromagnetic 
superexchange coupling between the $Fe^{2+}$ ions.  This picture was 
only fairly consistent 
with neutron diffraction results,$^{2}$ that indicated 
$\mu_{Fe} = 4.1(1)\mu_{B}$ and $\mu_{Mo}$ between $0.0(1)\mu_{B}$ 
and $0.42(6)\mu_{B}$.  Since these values were 
intermediate between the predictions of both ionic descriptions.  
M\"{o}ssbauer experiments$^{3}$ have been interpreted in terms of an intermediate 
valence of $m \cong 2.6$  
for the $Fe^{m+}$ ions. Recent X-ray absorption spectroscopy (XAS) experiments$^{4}$ have provided site-specific direct information on this problem: Fe is in the formal trivalent state and the magnetic moment at the Mo sites is negligible.  
These results are in 
agreement with band structure calculations$^{1,4}$ that predict that well localized 
spin up $t_{2g}$ and $e_{g}$ $Fe$ subbands are fully occupied.  The remaining 
electron goes into a spin down $t_{2g}$ delocalized subband formed by 
hybridized $Mo(4d)$ and $Fe(3d)$ orbitals, responsible for the metallicity of the material.  A well defined AFM
interaction results between the itinerant electrons  and 
the $Fe$ localized cores, driven by the hopping of the electrons between $Fe$ 
and $Mo$ sites.$^{4}$

Magnetic measurements in the paramagnetic (PM) phase 
should be able to provide useful evidence in order 
to establish the $Fe$ and $Mo$ valence in this compound and the 
possible interaction mechanisms.  Niebieskikwiat et al.$^{5}$ found that the 
high temperature magnetization, M, displays a 
non-conventional behaviour, interpreted in terms of two 
contributions arising from localized ($\mu_{eff} = 6.7 \mu_{B}/f.u.$) and itinerant electrons, respectively.  Preliminary measurements$^{6}$ in samples obtained under different thermal treatments have shown, for different applied fields, 
apparent values for $\mu_{eff}$ varying from $5.9\mu_{B}/f.u.$ to 
$4.5\mu_{B}/f.u.$ and suggested that this behavior was non intrinsic and due to the existence of antisite (AS) defects and to the presence 
of a ferromagnetic (FM) impurity.

Electron Spin Resonance (ESR) experiments also help 
to understand the magnetic properties of these perovskites.  The $g$ value
 brings information on the electronic 
structure of the ground state of the resonant ions and the linewidth is an 
experimental probe of the spin dynamics.  Niebieskikwiat et al.$^{5}$ observed 
a single ESR line whose intensity seemed to depart from a CW behavior and this result was 
interpreted in terms of a progressive delocalization of the $Fe^{3+}$ 
electrons.

In this paper we present detailed $dc$ magnetization and ESR measurements 
in the PM regime performed on a sample which presents an extremely low antisite defect concentration ($AS \cong 0.03$ and $M_{S} = 3.7$ $\mu_{B}/f.u.$) and only a 
small amount of a FM impurity phase ($\leq 0.5\%$).  This low AS value, determined by X-ray diffraction, was obtained by a careful control of the synthesis conditions (thermal treatment at $1200^{\circ}C$ for 12 hs in $5\%$ $H_{2}/Ar$), as described in Ref. 3.  Preparation of Fe impurities
 free samples is not an easy task since the very stable $SrMoO_{4}$ phase is readily formed above $800$ K under the presence of however small $O_{2}$ traces in the processing or measuring atmosphere.$^{7}$ and thus, severe reducing conditions are required.

\smallskip

We have measured M(T) vs H for $300$ K $\leq T \leq 1100$ K and for $H \leq 12.5$ kG, 
with a Faraday Balance Magnetometer.  The measurements were made in air 
($p < 1 torr$).  In order to control the reversibility, particularly in the high temperature range, we 
increased $T$ in $20$ K steps, repeating the measurements at $473$ K after 
each step.  By following this procedure we found irreversibles changes for $T > 800$ K.  Thus, we considered reliable only the
measurements in the range $300$ K-$800$ K and we analyze these data.  
The ESR experiments were performed with a Bruker spectrometer operating at 
$9.5 GHz$ between $300$K and $600$K.   

In Fig. 1 we show isotherms $M vs. H$.  For $T > 450$ K and high magnetic fields, we observed a 
linear dependence: $M(T) = M_{0}(T) + \chi(T)H$.  The parameters obtained 
for $H > 5$ kG, are given in Fig. 2.  The high field differential 
susceptibility, $\chi(T)$ follows a CW law for $T > 500$ K and up to $800$ K, the 
limit of the reversible behavior.  The fast increase of $ M_{0}(T)$ at low T indicates the 
FM transition at $T_{c} = 400$ K, determined from an Arrott plot 
(see inset in Fig. 1).  We note the presence of only a small ferromagnetic component well above $T_{c}$.  This contribution is weakly T dependent and varies 
between $M_{0}$($500$ K) $= 0.021$ $\mu_{B}/f.u.$ and 
$M_{0}$($800$ K) $= 0.013$ $\mu_{B}/f.u.$.  
This result is compatible with the presence of tiny amounts of 
$Fe$ impurities, as observed by X-ray Photoelectron Spectroscopy in epitaxial layers$^{8}$ of $Sr_{2}FeMoO_{6}$.

The ESR spectrum consists of a single line with lorentzian shape 
and $g = 2.01(2)$ for $T \geq 430$ K as described in a preliminary 
report$^{9}$.  
Above $450$ K the line broadens rapidly and we show in Fig. 3 
the peak-to-peak linewidth, $\Delta H_{pp}(T)$.  
The relative double integrated intensity of the line, $I_{ESR}$, decreases 
with increasing temperature, as shown in the inset.

Since
the spectrum broadens rapidly with increasing $T$, it is important to 
separate the contribution of the impurities, observed at high 
temperatures as a $T$ independent secondary line.$^{9}$  In order to obtain accurate values for $\Delta H_{pp}(T)$ and $I_{ESR}(T)$ of the principal line, we have separated the two contributions, for all temperatures above $480$ K, by subtracting the impurity spectrum which is almost temperature independent ($M_{0}$ varies less than $5\%$ between 480 K and 550 K) and the only visible at $T > 550$ K.

Important points to elucidate are the magnetic moment of the $Fe$ ions 
and the possible existence of a measurable Pauli contribution to the PM 
susceptibility.  We have determined separately 
at each temperature the FM contribution, $M_{0}(T)$, and the $\emph{true}$ 
PM susceptibility, $\chi(T)$, see Fig. 2.  We have found that 
$\chi(T)$ can be unambiguously described with a CW law, $\chi(T) = C/(T-\Theta)$, 
with $C = 2.68(9)$ emuK/mol and $\Theta = 446(3)$ K.  Any temperature independent 
contribution to $\chi(T)$ associated with the itinerant electrons was smaller that 
the experimental uncertainty ($\leq 5$ $10^{-4}emu/mol$).
The effective moment derived from the Curie constant should be corrected for the 
existence of AS defects.  For low AS concentration, an estimation of this effect may be obtained in a simplified mean field model, where misplaced Fe ions are strongly AFM coupled to their neighbors in regular sites.  The strength of this interaction is approximated by assuming the same coupling constant$^{10}$ as in the $LaFeO_{3}$ perovskite ($T_{N} = 750$ K).  This results in the formation of antiferromagnetic clusters around the AS defects, well above $T_{c}$, that reduce the average effective moment of the sample.  In our case the correction is $\approx 2\%$ and $\mu_{eff} = 4.72$ $\mu_{B}/f.u.$.

It should be emphasized that the value obtained for $\mu_{eff}$ is
smaller than expected for either $Fe^{3+}$ 
($\mu_{eff}= 5.9$ $\mu_{B}$) or $Fe^{2+}$ ($\mu_{eff} = 4.9$ $\mu_{B}$) ions.
Thus, the ionic picture associated with the description of $\chi(T)$ only in 
terms of fully localized $Fe^{3+}$ ( or even $Fe^{2+}$) moments can not account 
for 
the experimental results.  At this point we should ask what may be 
the contribution of the delocalized electrons to $\chi(T)$, 
taking into account their strong coupling with the $Fe$ 
cores, indicated by the band structure calculations.$^{1,4}$  An appropriate model for this situation was proposed by 
Vonsovky and Zener,$^{11}$ 
 describing the system in terms of localized ($M_{S}$) and mobile 
($M_{e}$) electrons in a mean field approximation.

\begin{equation}
	\mathbf{M}_{e} = \chi_{e}^{0}(T) \mathbf{H}_{e}^{eff} = \chi_{e}^{0}(T)(\mathbf{H} 
	+ \lambda \mathbf{M}_{S})	
\end{equation}
\begin{equation}
	\mathbf{M}_{S} = \chi_{S}^{0}(T) \mathbf{H}_{S}^{eff}  
	= \chi_{S}^{0}(T)(\mathbf{H} 
	+ \lambda \mathbf{M}_{e} + \alpha \mathbf{M}_{S})
\end{equation}
where $\chi_{S}^{0}$ and $\chi_{e}^{0}$ are the bare susceptibilities, 
in the absence of interactions, and $H_{e}^{eff}$, $H_{S}^{eff}$ are the 
effective fields acting on both sublattices. 
The polarization of the conduction band, induced by the localized moments,
is at the heart of this coupling mechanism.
This coupling 
may be either FM or AFM and it is represented by $\lambda$. A possible second neighbor 
$Fe^{3+}$-$Fe^{3+}$ superexchange interaction is denoted by $\alpha$.  
Solving equations (1-2), the total susceptibility of the coupled system is given by $\chi (T) = \chi_{S}(T) + \chi_{e}(T)$, where 

\begin{equation}
	\chi_{S}(T) =\frac{\chi_{S}^{0}[1 + (\lambda - \alpha)\chi_{e}^{0}]}
	{1 - (\alpha \chi_{S}^{0} + \lambda^{2}\chi_{e}^{0}\chi_{S}^{0})}	
\end{equation}	
\begin{equation}
 	\chi_{e}(T) =\frac{\chi_{e}^{0}(1 + \lambda\chi_{S}^{0})}
	{1 - (\alpha \chi_{S}^{0} + \lambda^{2}\chi_{e}^{0}\chi_{S}^{0})}
\end{equation}

If both $\chi_{e}^{0}$ 
and $\chi_{S}^{0}$ were Curie-like, $\chi (T)$ would have a typical ferrimagnetic behavior.  However, one of the coupled systems is 
delocalized and, therefore, its bare susceptibility, $\chi_{e}^{0}$, is temperature independent.  In this case, the total susceptibility, for 
$\chi_{S}^{0} = C_{S}/T$ with $C_{S} = \mu_{S}^{2}N_{A}/3k_{B}$, may be 
written as
\begin{equation}
	\chi(T) = \chi_{S}(T) + \chi_{e}(T) = \chi_{e}^{0} + \frac{C'}{T - \Theta}
\end{equation}

Here, two terms can be identified.  The first, temperature independent, 
is equal to $\chi_{e}^{0}$.  The second one is CW-like, where the Curie constant is now
$C' = C_{S}(1 +\lambda\chi_{e}^{0})^{2}$, renormalized because of the $S$-$e$ 
coupling. The effective moment of the coupled system is then given 
by $\mu_{eff} = \mu_{S}(1 + \lambda\chi_{e}^{0})$.  Therefore, we note that a reduction 
of $\mu_{eff}$ is expected for $\lambda < 0$, due to the 
AFM coupling of the itinerant electrons to the localized $Fe$ cores.  
The Curie Weiss temperature in Eq. (5) is given by 
$\Theta = C_{s}(\lambda^{2}\chi_{e}^{0} + \alpha)$ and
describes an effective interaction between the $S$ moments mediated by the 
delocalized electrons.  
Independently of the sign of $\lambda$, and for small $\alpha$, it is \emph{always} FM.  
In the double perovskite structure, 
$\alpha$ would be originated in superexchange interactions between second neighbors 
$Fe$ ions and it is indeed expected to be small.

The behaviour predicted by Eq. (5) is consistent with our experimental results, provided that $\chi_{e}^{0}$ is below the experimental resolution.  Based on the band structure calculations, and the XAS results$^{4}$ we can safely assume a $3d^{5}$ 
configuration for the localized $Fe$ cores, and then $\mu_{S} = 5.9 \mu_{B}$.  
From the measured $\mu_{eff} = 4.72 \mu_{B}$ and 
$\Theta = 446 K$, we derive 
$\chi_{e}^{0} = 3.7$ $10^{-4}$ emu/mol and $\lambda = -540$ mol/emu.  
The value obtained in this way for $\chi_{e}^{0}$ is, then, fully compatible with 
our $dc$ susceptibility measurements.  It is interesting to compare 
$\chi_{e}^{0}$ with available information 
in order to test its significance. The corrections for 
the Landau diamagnetism to the Pauli susceptibility and the Stoner 
amplification parameter may be derived by comparison between measured 
and calculated values in other metallic perovskites, such as$^{12}$
$LaNiO_{3}$.  By doing this we can estimate a density of states
at the Fermi level, $N(\varepsilon_{F}) = 2.8$ states/eV-f.u. 
for $Sr_{2}FeMoO_{6}$. Interestingly enough this value compares
well with band structure calculations.
With respect to $\lambda$, its negative value 
confirms the \emph{antiferromagnetic} coupling between the $Fe$ cores and the 
delocalized electrons at variance with the double exchange (DE) mechanism where localized and itinerant spins tend to be parallel.  This result supports the novel mechanism, 
kinetically driven, described by Sarma et al.$^{4}$  It should be emphasized 
that, unlike typical ferrimagnets, the CW temperature is positive, 
in spite of the AFM character of the interaction.  Notice that 
$\Theta$, and consequently $T_{c}$, is proportional to $\chi_{e}^{0}$. Within this picture a larger density of states at the Fermi level should promote a higher $T_{c}$.

We can now turn to the ESR results.  We have found that 
$I_{ESR}(T)$ follows the same temperature 
behavior as $\chi(T)$, in the whole PM region (see inset Fig. 3).  
This observation 
indicates that the same magnetic species contributes to the ESR spectrum 
and the $dc$ susceptibility. 
Our ESR spectrum should also shed light on the issue of the valence 
of $Fe$ ions.  The measured gyromagnetic factor, $g = 2.01(1)$, is $T$ 
independent and may be identified with the spin-only ground 
state of $Fe^{3+}$ ions.  This resonance corresponds, 
in the band picture, to the localized $3d^{5}$ $Fe$ cores.  The observed g-value 
discards the possibility of 
assigning the resonance to localized $Fe^{2+}$ ($g \cong 3.4$).  Our $dc$ 
susceptibility measurement indicates a 
strong coupling between these localized $Fe$ cores 
and the itinerant electrons.  
The influence of this coupling on the spin dynamics is  
described by the Bloch-Hasegawa (BH) type equations.$^{13}$
\begin{equation}
\frac{d\mathbf{M}_{e}}{dt} = \frac{g_{e}\mu_{B}}{\hbar}(\mathbf{M}_{e} \times 
\mathbf{H}_{e}^{eff}) - \left(\frac{1}{T_{eL}}+ \frac{1}{T_{eS}}\right)
(\mathbf{M}_{e} - 
\chi_{e}^{0}\mathbf{H}_{e}^{eff})
+ \frac{g_{e}}{g_{S}}\frac{1}{T_{Se}}(\mathbf{M}_{S} - 
\chi_{S}^{0}\mathbf{H}_{S}^{eff})
\end{equation}

\begin{equation}				
\frac{d\mathbf{M}_{S}}{dt} = \frac{g_{S}\mu_{B}}{\hbar}(\mathbf{M}_{S} \times 
\mathbf{H}_{S}^{eff}) - \left(\frac{1}{T_{SL}} + \frac{1}{T_{Se}}\right)
(\mathbf{M}_{S} - 
\chi_{S}^{0}\mathbf{H}_{S}^{eff}) 
+ \frac{g_{S}}{g_{e}}\frac{1}{T_{eS}}(\mathbf{M}_{e} - 
\chi_{e}^{0}\mathbf{H}_{e}^{eff}) 	
\end{equation}
where $H_{e}^{eff}$ and  $H_{S}^{eff}$, defined by Eqs. (1) and (2), 
are now the instantaneous effective fields (including the $rf$ field).  
Here, 
$1/T_{eL}$ and $1/T_{SL}$ are the spin-lattice relaxation rates for 
delocalized and localized spins, respectively and 
$1/T_{Se}$, $1/T_{eS}$ the cross relaxation rates.  In our case values of 
$g_{S}$ and $g_{e}$ are both expected to be very close to $g \cong 2$.	

The 
solutions of Eqs. (6) and (7) present two well differentiated regimes: 
bottlenecked and non-bottlenecked, associated with the relative importance of 
the coupling between the equations.$^{13}$  In the non-bottlenecked case both 
systems tend to respond independently and two resonances should be observed, with $g$-shifts related to the corresponding 
effective fields.  In the bottleneck limit the strong coupling 
of Eqs. (6) and (7) results in a single resonance line corresponding to 
the response of the weighted sum 
$\mathbf{M} = \mathbf{M}_{e}/g_{e} + \mathbf{M}_{S}/g_{S}$ to the $rf$ field ($\emph{symmetric-mode}$).  
The other solution of the BH equations ($\emph{antisymmetric-mode}$) has no coupling to the 
$rf$ field and, therefore, does not contribute to 
the resonance spectrum.  The symmetric mode has an effective $g$-value
\begin{equation}
g = [g_{e}\chi_{e}(T) + g_{S}\chi_{S}(T)] / \chi(T)		
\end{equation}
where $\chi_{S}(T)$ and $\chi_{e}(T)$ were defined in Eqs. (3) and (4), 
respectively and the linewidth is given by  
\begin{equation}
\Delta H_{pp}(T) = \frac{2\hbar}{\sqrt{3}g\mu_{B}}
\left(\frac{\chi_{e}^{0}}{\chi (T)}\frac{1}{T_{eL}} + 
\frac{\chi_{S}^{0}(T)}{\chi (T)}\frac{1}{T_{SL}}\right) 		
\end{equation}

In our case, where $g_{S} \cong g_{e}$, $M$ is  the total magnetization 
of the system and a temperature independent $g \cong 2$ is obtained 
from Eq. (8).  
Since $\chi_{S}^{0}(T) = C_{S}/T \gg \chi_{e}^{0}$, 
in the whole $T$ range of our experiment, the linewidth 
may be approximated by 
$\Delta H_{pp}(T) \cong [C_{S}/(T\chi(T))]\Delta H_{pp}^{\infty}$, 
where $\Delta H_{pp}^{\infty}$, 
the linewidth in the high $T$ limit, is dominated 
by the relaxation rate of the localized cores, $1/T_{SL}$. 
 
We obtained a good fit of the 
data using a temperature independent $\Delta H_{pp}^{\infty} = 14(1) kG$, 
as 
seen in Fig. 3.  The relaxation rate, for strongly localized interacting 
spins results from the balance between \emph{broadening} 
(dipolar, antisymmetric exchange, crystal field), $\omega_{a}$, 
and \emph{narrowing} (isotropic exchange), $\omega_{e}$, interactions between the $Fe$ cores: 
$1/T_{SL} \propto (\omega_{a}^{2}/\omega_{e}$).  The value obtained here for 
$\Delta H_{pp}$ is very large, as compared with those found in other $Fe^{3+}$ 
oxides$^{10, 14}$ with ordering temperatures around $200 K$-$750 K$, where 
$\Delta H_{pp}^{\infty}$ varies between $0.5 kG$ and $1.7 kG$.  Since we do 
not expect large variations of $\omega_{a}$ in perovskite oxides, we assume 
that the larger linewidth must be due to a less important degree of 
exchange narrowing (small $\alpha$).

In CMR manganites, where the conventional double exchange mechanism is responsible for 
strong FM interactions, a similar behavior was observed: the increase in the 
ordering temperature is not accompanied by an enhancement of the exchange narrowing of the 
ESR line.$^{15}$  In the case of $Sr_{2}FeMoO_{6}$, the reason for this 
behavior can be rationalized in terms of the BH equations.  The ordering 
temperature is determined by the combined effect of the 
$S$-$e$ coupling ($\lambda$) and the $S$-$S$ superexchange ($\alpha$).  The 
narrowing of the linewidth, instead, depends only on $\alpha$.  Taking into 
account that in the $Fe$ compounds referred before,$^{10, 14}$ the 
$Fe^{3+}$ ions are nearest neighbors it is not surprising that the narrowing 
effect in $Sr_{2}FeMoO_{6}$ is smaller because $Fe^{3+}$ ions 
are second neighbors 
in this case.

In summary, we have obtained an excellent and consistent description of the experimental 
results of $dc$ susceptibility and ESR spectroscopy, in terms of a system of 
two coupled equations for the $Fe^{3+}$ localized cores (indicated by $g \cong 2$) and the itinerant 
electrons, delocalized in both $Fe$ and $Mo$ sites.  By solving these equations in a mean-field 
approximation we can account for the 
reduction of the effective moment of the $3d^{5}$ $Fe$ cores in the PM 
regime, due to the strong AFM coupling with the itinerant electrons.  The 
delocalized nature of these electrons, with a $T$ independent bare 
susceptibility, causes a non-conventional ferrimagnetic behavior, where 
$\Theta$ is positive and equal to $T_{c}$, in spite of the AFM character 
of the interaction.  This is in agreement with the picture presented by Sarma et 
al$^{4}$.  Due to the Pauli exclusion principle, the $t_{2g}$ electrons are 
allowed to hop from site to site only if the $Fe$ core spins are all 
oriented antiparallel to them.  As 
a consequence, a FM state with all $Fe$ spins parallel is energetically 
favored.  Within the present framework, the density of states of the
mobile electrons, being proportional to $\chi_{e}^{0}$, plays an important
role in the determination of the ordering temperature. The enhancement of 
$T_{c}$ found$^{16}$ when double perovskites are electron doped, nicely fits 
in this model.

Within the same picture we have described the spin dynamics of 
the strongly coupled system extending the use of the Bloch-Hasegawa 
equations to materials magnetically concentrated, where 
$\chi_{S}^{0}(T) \gg \chi_{e}^{0}$ even at high temperatures. In this 
case, the effective relaxation rate for the coupled system is 
dominated by $1/T_{SL}$, the relaxation rate of the localized spins.  
The absence of narrowing effects associated with the 
high $T_{c}$ is consistent with
the fact that the dominant 
mechanism for magnetic ordering in $Sr_{2}FeMoO_{6}$ 
is a process where the FM coupling between Fe ions is mediated by the mobile 
electrons.  

We thank Dr. B. Alascio for helpful comments.  This work
was partially supported by: CONICET Argentina (PIP 4749), ANPCyT 
Argentina (PICT 03-05266) and the projects: AMORE from the CEE and MAT 1999-0984-CO3 from the CICyT Spain.

\begin{figure}
FIG. 1.  $M$ $vs.$ $H$ at different temperatures.  
Inset: Arrott's plots ($M^{2} vs. H/M$) around $T_{c}$.
\end{figure}

\begin{figure}
FIG. 2.  $\chi^{-1}$ $vs.$ $T$.  Inset: $M_{0}$ $vs.$ $T$.
\end{figure}

\begin{figure}
FIG. 3.  $\Delta H_{pp}(T)$ $vs.$ $T$.  The line indicates 
the fitting with $\Delta H_{pp}^{\infty} = 14(1) kG$.  Inset: $I_{ESR}(T)$ \emph{vs.} $T$, dotted line corresponds to $\chi (T)$. 
\end{figure}

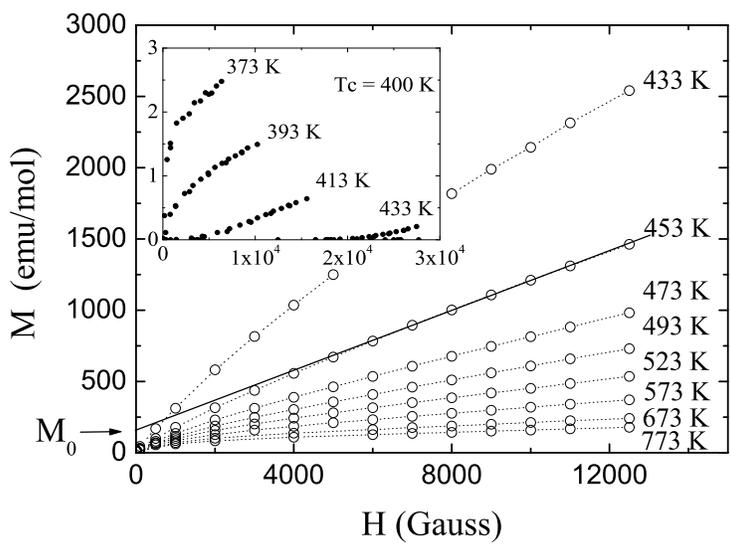
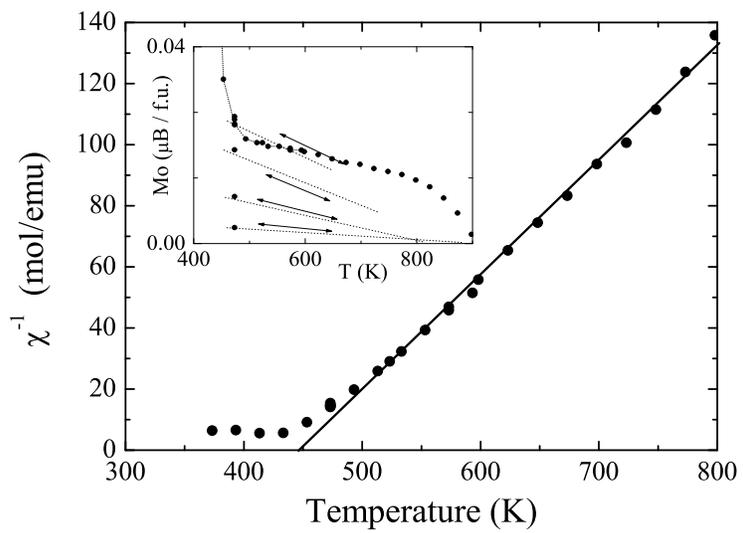
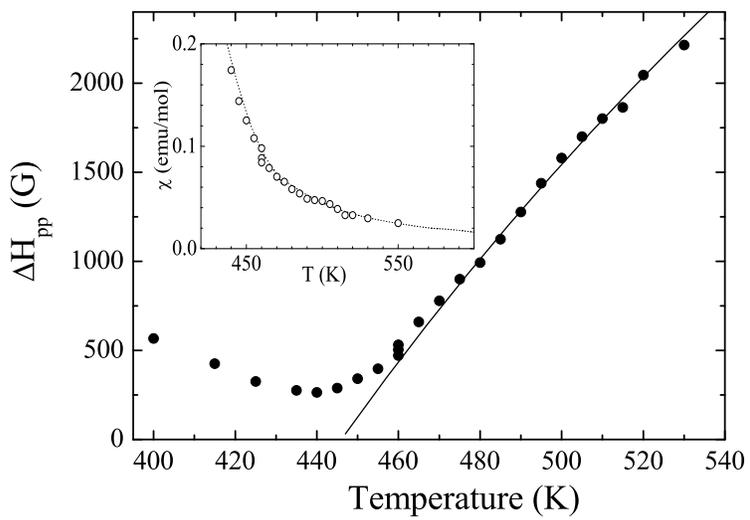
\end{document}